\documentstyle{article}

\begin{document}

\title{Compositeness effects, Pauli's principle and entanglement}

\author{Pedro Sancho \\ GPV de Valladolid \\ Centro Zonal en
Castilla y Le\'on \\ Ori\'on 1, 47014, Valladolid, Spain}
\maketitle

\begin{abstract}
We analyse some compositeness effects and their relation with
entanglement. We show that the purity of the composite system
increases, in the sense of the expectation values of the deviation
operators, with large values of the entanglement between the
components of the system. We also study the validity of Pauli's
principle in composite systems. It is valid within the limits of
application of the approach presented here. We also present an
example of two identical fermions, one of them entangled with a
distinguishable particle, where the exclusion principle cannot be
applied. This result can be important in the description of open
systems.
\end{abstract}

\vspace{1mm}


\vspace{1mm}

\section{Introduction}

According to the usual textbook rule, when in a physical process
the internal structure of a composite particle is not revealed,
the particle should approximately behave as a boson or a fermion
depending on the number of constituent fermions and bosons. When
the internal structure of the particle is taken into account,
compositeness effects can appear which manifest in deviations from
the purely bosonic or fermionic behaviour. These deviations have
been studied in several contexts. From a fundamental point of
view, the interactions between composite bosons have been analysed
highlighting the differences with the case of pure bosons
\cite{Comb,Wojs}. On the other hand, several systems where these
deviations can take place have been studied from both the
theoretical and experimental points of view. There are two principal lines, Bose-Einstein condensation (BEC) and
semiconductors. The BEC occurs even when the atoms of a condensate
gas are not pure bosons. Small deviations of the pure behaviour in
a BEC have been presented in \cite{bra}. In the case of
semiconductors, the research has focused on excitons
(electron-hole bound pairs). Criteria for the bosonic behaviour of
excitons have been derived in \cite{Com}. From an experimental
point of view, the excitons can be studied by optical spectroscopy
(photoluminescence, reflectivity, etc). In particular, there has
been an extensive analysis of the optical spectra of quantum wells
(two-dimensional structures containing electron gases \cite{prb}).
We must also mention the issue where both lines, BEC and
semiconductors, converge, the BEC of excitons. It was predicted by
Keldysh and Zozlov \cite{rusos}, being quantum degeneracy
experimentally observed in the system later \cite{But}.

More recently, the close relation between
entanglement and the purity of a composite boson has been signaled \cite{Law}. That
analysis was based on the properties of the creation and
annihilation operators of the composite boson. In this paper, we
show with a particular example that a similar result is
obtained when the purity of the composite boson is measured in
terms of the expectation values of the deviation operators.

In addition, and as the main aim of the paper, we study in a
rigorous way the validity of Pauli's principle in "composite
fermions" of the type considered in this paper. To our knowledge,
this issue has not previously been considered in the literature.
We shall show that in the range where the theory considered here
is applicable the exclusion principle remains exactly valid for
composite particles. On the other hand, we shall present an
example where if one of two identical fermions in the same state
becomes entangled with a third distinguishable particle  the
exclusion principle does no longer act between them. This result
is potentially important in open systems, where the fermions of the
pair can interact with a large number of particles. If these
interactions are of the type that lead to entanglement the
exclusion principle can become fragile. This result shows striking
resemblances with decoherence theory \cite{Zeh}.

We shall restrict our considerations to systems composed of two
particles in multimode states. The starting point of the analysis
will be the (anti)commutation relations of creation and annihilation
operators. We shall present the set of
(anti)commutation relations in a general way, including those of
composite particles in different multimode states.

The plan of the paper is as follows. In Sect. 2 we present the
general set of (anti)commutation relations. Section 3 deals with
an analytic example of the evaluation of the expectation values of
the deviation operators. In Sects. 4 and 5 we study, respectively,
the validity of the exclusion principle in composite systems and
in the presence of entanglement with other particles. In Sect. 6 we emphasize on the main results of the paper. Finally,
in the Appendix we present an alternative derivation of the
results of Sect. 5.

\section{The (anti)commutation relations}

The wavefunction of a system composed of two distinguishable
particles in continuous multimode states is given by
\begin{equation}
\psi (x,y)=\int \int dpdq f(p,q) \Psi _{p,r} (x) \phi _{q,s} (y)
\label{eq:uno}
\end{equation}
where $x$ and $y$ are the coordinates associated with the two
particles (by simplicity we only consider the one-dimensional
problem). $\Psi _{p,r}$ and $\phi _{q,s}$ are the wavefunctions
corresponding to modes with momentum $p$ and $q$ in spin states
$r$ and $s$ (the same spin state for all the modes of every particle). On the other
hand $f(p,q)$ is the distribution of
modes. We assume by simplicity that the distribution of modes is
independent of the spin states. The integrations in Eq.
(\ref{eq:uno}) extend between $-\infty$ and $\infty$ (just as all the
other integrals appearing in the paper).

In the second quantization formalism, the composite particle represented by Eq. (\ref{eq:uno})
corresponds to the
state generated by the creation operator
\begin{equation}
\hat{c}_f^+ = \int \int dpdq f(p,q)\hat{a}_{p,r}^+ \hat{b}_{q,s}^+
\end{equation}
where $\hat{a}_{p,r}^+$ and $\hat{b}_{q,s}^+$ are the creation operators of modes
$\Psi _{p,r}$ and $\phi _{q,s}$ (as signalled before we assume both particles to be
distinguishable ones).

After simple calculations using the relations $[ \hat{a}_{p,r} ,
\hat{a}_{P,s}^+ ]=\delta (p-P) \delta _{rs}$,
 $[ \hat{a}_{p,r}^+ , \hat{a}_{P,s}^+ ]=0$ and $[ \hat{a}_{p,r} , \hat{a}_{P,s} ]=0$ for bosons
and $\{ \hat{a}_{p,r} , \hat{a}_{P,s}^+ \}=\delta (p-P) \delta
_{rs} $, $\{ \hat{a}_{p,r}^+ , \hat{a}_{P,s}^+ \}=0$ and $\{
\hat{a}_{p,r} , \hat{a}_{P,s} \}=0$ for fermions, the
(anti)commutation relations between the creation and annihilation
operators of composed particles with different mode distributions
$f$ and $g$ are as follows (we denote the spin states of particles $a$
and $b$ in mode $f$ by $r$ and $s$, and those in mode $g$ by $R$
and $S$, respectively):
\begin{equation}
[ \hat{c}_f , \hat{c}_g ]_{BB}= [ \hat{c}_f , \hat{c}_g ]_{FF}= \{
\hat{c}_f , \hat{c}_g \}_{FB} =0
\end{equation}
\begin{equation}
[ \hat{c}_f^+ , \hat{c}_g^+ ]_{BB}= [ \hat{c}_f^+ , \hat{c}_g^+
]_{FF}= \{ \hat{c}_f^+ , \hat{c}_g^+ \} _{FB} =0
\label{eq:cuat}
\end{equation}
\begin{equation}
[ \hat{c}_f , \hat{c}_g^+ ]_{BB}= \theta \hat{1} + \hat{\theta }_a
+\hat{\theta }_b
\end{equation}
\begin{equation}
[ \hat{c}_f , \hat{c}_g^+ ]_{FF}= \theta \hat{1} - \hat{\theta }_a -
\hat{\theta }_b
\end{equation}
and
\begin{equation}
\{ \hat{c}_f , \hat{c}_g^+ \} _{FB}= \theta \hat{1} - \hat{\theta }_a +
\hat{\theta }_b
\end{equation}
with
\begin{equation}
\theta =\int \int dp dq f^*(p,q)g(p,q)\delta _{rR} \delta _{sS} =<f|g> \delta _{rR} \delta _{sS}
\end{equation}
\begin{equation}
\hat{ \theta}_a =\int \int \int dp dP dq f^*(p,q)g(P,q)\hat{a}_{P,R}^+ \hat{a}_{p,r}
\delta _{sS}
\end{equation}
and
\begin{equation}
\hat{ \theta}_b =\int \int \int dp dq dQ f^*(p,q)g(p,Q)\hat{b}_{Q,S}^+ \hat{b}_{q,s}
\delta _{rR}
\end{equation}
In the above expressions symbols $[ , ]$ and $\{ , \}$ refer,
respectively, to commutators and anticommutators. On the other
hand, $BB$, $FF$ and $FB$ denote particles composed of two bosons,
two fermions or a fermion and a boson. In the case of
boson-fermion systems we adopt the convention that $a$ refers to
the fermion and $b$ to the boson. Variable $\theta $ has been
expressed in terms of $<f|g>$, the scalar product of wavefunctions $f$ and $g$ in momentum space. This expression shows
that when $f$ is orthogonal to $g$, $\theta$ vanishes.

We obtain a set of relations that differs from the pure bosonic
and fermionic algebras. Only when $\theta =1$, $\hat{ \theta}_a
=0$ and $\hat{ \theta}_b=0$ can we recover the usual (anti)commutation
relations for bosons and fermions. This property resembles that of
the quon algebra, which interpolates between the bosonic and
fermionic ones \cite{var} (it has been used in Ref. \cite{bra} to
study compositeness effects).

We note that these commutation relations are
different for boson-boson (BB) and fermion-fermion (FF) systems.

It is also simple to see that $\hat{ \theta}_a $ and $\hat{
\theta}_b$ are not, in general, nulls, even in the absence of
common modes. Consequently, the (anti)commutator of
the creation and annihilation operators with different
distributions can be different from zero even in the case when
they have no common modes. This result does not
follow the intuition obtained in the study of systems
in multimode states. For instance, in Ref. \cite{San} it was shown
that the existence of common modes is  a necessary
condition for the existence of interference effects in the
arrangement considered in that paper. This property could, in
principle, be experimentally tested in systems whose interactions
depend on the (anti)commutator of the creation and annihilation
operators.

\section{Expectation values of deviation operators}

One of the ways of characterizing the importance of the new
effects in composite systems is via the expectation values of the
deviation operators. We say that a system behaves as a pure boson
(fermion) when their creation/annihilation operators obey the Bose
(Fermi) commutation (anticommutation) relations. When a composite
system does not obey these relations the departure from the pure
behaviour can be quantitatively estimated by some measure of the
difference between the relations of pure and composite systems. As
remarked in the previous section the operators $\theta \hat{1}$
(different from $\hat{1}$), $\hat{\theta }_a$ and $\hat{\theta }_
b$ (different from $0$) are responsible for the departures from
the usual algebras, and can be named deviation operators.
The importance of the deviation
operators depends on the state of the system. We must evaluate
their expectation values on the state of the system. We use these
expectation values as a measure of the degree of purity of the
system (how close the system is to a pure one).

In order to illustrate the general method, we shall evaluate in
this section these expectation values for two identical "composite
bosons". The distributions and spin states must also be equal
($f=g$, $r=R$ and $s=S$), and are no longer necessary to include
explicitly the spin indexes in the expressions. The state of the
system is $|2_f >= N(c_f^+)^2|0>$, with $N$ the normalization
factor, given by
\begin{equation}
2 N_{   _{ FF} ^{BB}  }^2 =\frac{1}{1 \pm \wedge }
\end{equation}
where
\begin{equation}
\wedge = \int \int \int \int dpdPdqdQ f^*(P,Q)f(P,q)f^*(p,q)f(p,Q)
\end{equation}

In this expression and from now on, in all the
double expressions the upper sign corresponds to the $BB$ case and
the lower one to the $FF$ one.

To obtain the above equations we have used the
normalization of the distributions, $\int \int dpdq |f(p,q)|^2=1$.
This normalization emerges directly from the usual normalization
of the wavefunction, $\int \int dxdy |\psi (x,y)|^2=1$, and the
orthogonality relations between the $\Psi$ and $\phi$.

In this case, since both particles are equal we have $f=g$ and
using the normalization relation we obtain $\theta =1$. The
expectation values are:
\begin{eqnarray}
<2_f| \hat{\theta }_a |2_f> _{ _{FF}^{BB}} = <2_f| \hat{\theta }_b |2_f> _{ _{FF}^{BB}} = N_{ _{FF}^{BB}}^2 \times \nonumber \\
\int \int \int \int \int \int dp dP dp_* dq dQ dq_* F(p,..,q_*)G(p,..,q_*)
\end{eqnarray}
where
\begin{equation}
F(p,..,q_*)=f^*(p,q)f^*(P,Q)f^*(p_*,q_*)
\end{equation}
and
\begin{eqnarray}
G(p,..,q_*)= 2f(p_*,Q)f(P,q_*)f(p,q)+ 2f(p_*,q)f(p,q_*)f(P,Q) \pm  \nonumber \\
 2f(p_*,q)f(p,Q)f(P,q_*) \pm  2f(p_*,Q)f(P,q)f(p,q_*)
\end{eqnarray}
To illustrate the behaviour of the above expressions we choose a
tractable example of distribution:
\begin{equation}
f(p,q)= \sqrt{\frac{2}{\pi}}(\alpha \beta - \gamma ^2)^{1/4} exp(-\alpha p^2 - \beta q^2 -2
\gamma pq)
\label{eq:nene}
\end{equation}
with $\alpha >0$ and $\beta >0$, both real for simplicity. Taking
$\alpha \beta \geq \gamma ^2$ this distribution is a Gaussian one.
We note that the distribution is normalized. In the absence of the
third term in the exponential, $\alpha ^2 $ and $\beta ^2 $ are
the widths along directions $p$ and $q$, measuring the spread of
the distribution. When $\gamma =0$ the distribution can be
separated into two independent distributions of $p$ and $q$. If
$\gamma \neq 0$ the distribution is no longer separable. As the
distribution $f$ is the wavefunction in momentum space, $\gamma
\neq 0$ can be associated with the presence of entanglement in the
system. This can be directly verified by transforming to the
position space where, for $\alpha =\beta $, we have $\psi (x,y)
\sim exp(-(\alpha x^2 +\alpha y^2 +2\gamma xy)/4(\alpha ^2 -
\gamma ^2)\hbar ^2)$. We see again that for $\gamma \neq 0$ the
wavefunction cannot be separated. A measure of the degree of
entanglement is $\gamma ^2 /\alpha \beta $. For $\gamma =0$ the
entanglement vanishes. On the other hand, for $\gamma ^2
\rightarrow \alpha \beta $ the expression tends towards the
maximum value, $1$, which for $\alpha = \beta $ corresponds to
$\gamma \rightarrow \pm \alpha $, which gives $\alpha (p^2 +q^2)
\pm 2\gamma pq \rightarrow \alpha (p \pm q)^2$. The distribution
has only appreciable values for $p \rightarrow \pm q$ ($|p|
\approx |q| $).

For this particular distribution, the normalization and expectation values become:
\begin{equation}
2 N_{ _{FF}^{BB}  }^2 =\frac{1}{1 \pm \epsilon }
\label{eq:dise}
\end{equation}
and
\begin{eqnarray}
<2_f| \hat{\theta }_a |2_f> _{ _{FF}^{BB}} = <2_f| \hat{\theta }_b |2_f> _{ _{FF}^{BB}} =
\frac{2 \epsilon }{1 \pm \epsilon } \pm \nonumber \\
\frac{8 \epsilon ^3 (\alpha \beta )^{3/2}}{ \sqrt{2 \mu \eta }
(2\alpha \beta - \gamma ^2)^{1/2} ( 1 \pm  \epsilon ) }
\label{eq:disi}
\end{eqnarray}
with
\begin{equation}
\epsilon = \left( 1-\frac{\gamma ^2}{\alpha \beta }  \right)
^{1/2}
\end{equation}
\begin{equation}
\mu = 2 \alpha \beta - \gamma ^2 - \frac{\gamma ^4}{4(2\alpha \beta -\gamma ^2 )}
\end{equation}
and
\begin{equation}
\eta = \mu - \frac{1}{4\mu }     \left( \gamma ^2 +  \frac{\gamma ^4
}{2(2\alpha \beta -\gamma ^2 )} \right) ^2
\end{equation}

The new parameter $\epsilon $ is the square root of
$1-\frac{\gamma ^2}{\alpha \beta }$. As $\frac{\gamma ^2}{\alpha
\beta }$ is a measure of the entanglement of the system $\epsilon
$ will be a measure of its complementary variable, i. e., of the
degree of separation of the state. When $\gamma =0$, $\epsilon =1$
and the state can be completely separated. On the other hand, when
$\gamma ^2 =\alpha \beta $, $\epsilon =0$ reaching its minimum
separation.

In order to carry out these integrals with the expression $\int dx
exp(ax^2 +bx)= (-\pi /a)^{1/2} exp(-b^2/4a)$, $a$ must obey the
relation $a \leq 0$. In the case of Eq. (\ref{eq:dise}) this
relation gives:
\begin{equation}
2\alpha - \frac{\gamma ^2}{\beta } \geq 0  \; ; \; 2\alpha -
\frac{\gamma ^2}{\beta } - \frac{\gamma ^4}{\beta (2 \alpha \beta
- \gamma ^2)}\geq 0
\end{equation}
It is simple to see that these relations automatically hold when
$\alpha \beta \geq \gamma ^2$. In the case of Eq. (\ref{eq:disi})
the relations are
\begin{equation}
2\alpha - \frac{\gamma ^2}{\beta } \geq 0  \; ; \; \frac{\mu
}{\beta } \geq 0 \; ; \; \frac{\mu }{\beta } -\frac{\left( \gamma
^2 + \frac{\gamma ^4}{2(2 \alpha \beta - \gamma ^2)} \right)  ^2}
{4\beta \mu  } \geq 0
\end{equation}
As in the previous case it is simple to show after some
manipulations that all these relations are valid when $\alpha
\beta \geq \gamma ^2$

When $\gamma ^2 \rightarrow \alpha \beta $ (the maximum
entanglement of the system) the
expectation values of the deviation operators tend to zero for
both BB and FF systems. The composite systems behave, in the sense
of expectation values, as pure bosonic systems.
On the other hand, when
$\gamma ^2 \neq \alpha \beta $ the expectation values are, in
general, different from zero.

This result agrees with those of Ref. \cite{Law}, where it was
demonstrated that bosonic character emerges when the constituent
particles become strongly entangled. It must be remarked that in
that reference the results were derived studying the properties of
the creation and annihilation operators of the composite
particles. In this paper, we have based our analysis on the
expectation values of the deviation operators, providing an
independent confirmation of the connection between almost pure
bosonic behaviour and strong entanglement.

It must also be remarked that the expectation values are different
for FF and BB systems because of the different signs in the
expressions of type $\pm $. This property is specially
striking for completely decorrelated particles ($\gamma =0$). This
limit will be considered in the next section.

\section{Pauli's principle in composite systems}

We devote Sects. 4 and 5 to the analysis of Pauli's principle in
composite systems. First, in this section, we study the behaviour
of identical "composite fermions".

An important consequence concerning Pauli's principle follows
directly from the Relations derived in Sect. 2. From the
anticommutation relations in Eq. (\ref{eq:cuat}) we have
$(\hat{c}_f^+)^2=0$, which is the mathematical expression of the
exclusion principle. Effectively, if we try to create a state with
two identical "composite fermions" we must apply the operator
$(\hat{c}_f^+)^2$ on the vacuum $|0>$ with the result
$(\hat{c}_f^+)^2 |0>=0$. It is impossible to prepare such a state.
In addition to the case of systems composed of two particles, it is
simple to verify by direct calculation that this property is valid
for any type of particle composed of an odd number of
(distinguishable) fermions. We conclude that it is impossible to
prepare two or more "composite fermions" of the type considered here in the same state. Where the results presented here are
valid they provide a justification for the use of the principle in
non-pure conditions, although the complete set of anticommutation
relations differs from the pure one (we do not have the pure
fermion algebra).

Of course, the validity of this result is limited by the scope of
the framework considered here. Essentially, this limit is given by
the validity of the description of the two-particle system as a
single entity in the second quantization formalism. Although mathematically this
procedure can always be carried out it can be misleading from a physical point of view.

In order to clarify this point let us consider an example. The
second quantization formalism is relevant for the problem when the
creation/annihilation operators represent correctly the physical
processes involved. Let us consider, for instance, the
annihilation operator and a situation where a composite particle
interacts with an absorbing medium, i. e., with one that can
capture and absorb particles (the composite one and/or any of its
two components $a$ and $b$). If, for instance, the medium can
absorb particle $a$ without absorbing $b$ (and, consequently,
without absorbing the composite system) the annihilation operator
of particle $a$ is relevant for the problem, but not that of the
composite system. In this case, the description of the composite
system in the second quantization formalism as a single entity
represented by only an annihilation operator is meaningless. For
bound systems with the two particles $a$ and $b$ close one
expects, in general, the operator of the composite system to be
relevant. However, for entangled separated systems we have no {\em a
priori} general criterion for this problem. The analysis of the
relevance of the annihilation (or creation) operator must be
carried out for every system.

\section{Pauli's principle and entanglement}

A detailed analysis of the example presented in Sect. 3 in the
limit of no correlation in the mode distributions shows an
interesting connection between entanglement and Pauli's principle.
We devote this section to study this connection. We start our
discussion by considering that example.

\subsection{An example}

We see that for FF systems expression (\ref{eq:dise}) becomes
unbounded (and then physically forbidden) when the distribution
can be factored, i. e., when $\gamma =0$ and the distributions of
the two particles are uncorrelated. Moreover, Eq. (\ref{eq:disi})
shows that when $\gamma =0$, the expectation values of the
deviation operators of composed FF systems become of the
undetermined form $0/0$. These results reflect the impossibility

of preparing the system in such a state. No counterpart to such
critical behaviour is found for BB systems, for which a finite
value is reached.

These results are manifestations of Pauli's principle, which acts
in the absence of entanglement ($\gamma =0$). When the particles
are not entangled, we have two free indistinguishable fermions of
every class ($a$ and $b$). Pauli's principle acts between them.
However, when there is entanglement the exclusion principle does
no longer act. We find again a close relationship between the
behaviour of composite systems and entanglement.

This is not only a property of the distribution considered here. For any pair
of composite particles of the FF type in a non-entangled state the
distribution $f(p,q)$ is separable in the form $f_a(p)f_b(q)$.
Then using the normalization condition of the distributions we
have $\wedge =1$. The normalization constant is unbounded
reflecting the impossibility of preparing the state (as a
consequence of Pauli's principle). However, if $f(p,q)$ is
non-separable, in general, $\wedge \neq 1$ and the normalization
constant is finite.

\subsection{Two fermions, one in an entangled state}

These results are only an example of a more general and important
property. Let us imagine two identical non-composite fermions.
Initially they are very separated (then the overlapping between
their wavefunctions is negligible and it is not necessary to
antisymmetrize the complete wavefunction \cite{Mes}). One of them
can interact with a third particle of a different type. The
interaction will result in many cases in an entangled state of the
fermion and the third particle (the interaction Hamiltonian can in
many cases couple both particles). Let us consider the case when
this pair of entangled particles must be described as a single
entity in the second quantization formalism (see the discussion at
the end of the previous section). Later, the two fermions come
together. A particular realization of this scheme (of a Gedanken
type) is as follows. The distinguishable particle and one of the
indistinguishable ones are enclosed in a box where they interact and
become entangled. On the other hand, the other indistinguishable
particle is placed in another box. Both boxes are separated by a

common movable wall, which prevents any overlapping between the
identical particles (as remarked before, without overlapping it is
not necessary to take into account the indistinguishable character of the particles \cite{Mes}).
Later we remove the wall and the overlapping of the
indistinguishable particles becomes appreciable. If necessary,
before removing the wall we displace the distinguishable particle
in order for the interaction with the indistinguishable particle
(that previously in the other box) to be negligible.

We assume that the state of the free fermion and the
local state (obtained by local measurements on the fermion)
of the entangled fermion are the same. Note that an observer can
be unaware of the interaction with the third particle,
consequently identifying the local state of the entangled fermion
with the state of the fermion. We study if the exclusion principle
can be applied in this case.

The three-particle state is $|3>=N_3 \hat{c}_f^+ \hat{a}_g^+ |0>$ with
$\hat{c}_f^+$ and $\hat{a}_g^+$ the creation operators of the
composite particle and the free fermion, respectively. The normalization constant is given by:
\begin{eqnarray}
N_3^{-2} = \int \int \int dp dP dq (g^*(p)f^*(P,q)g(P)f(p,q)- \nonumber \\
g^*(p)f^*(P,q)g(p)f(P,q))=  \\
-1+ \int \int \int dp dP dq g^*(p)f^*(P,q)g(P)f(p,q) \nonumber
\end{eqnarray}

When there is no interaction between the fermion and the third
particle or the interaction does not induce entanglement $f(p,q)$
can be factored in the form $g(p)F(q)$ (with the two identical
fermions having the same mode distribution $g(p)$). $N_3$ becomes
unbounded and the state of the system cannot be normalized.
Without normalization the state cannot be interpreted along the
usual statistical formulation of quantum theory. Consequently, it

is not a physical state that can be associated with the system in
the framework of standard quantum theory. In other words, it is not a state
that can be prepared following the usual schemes. Pauli's
principle acts between the two fermions, avoiding the possibility
of preparing the three particles in that state. On the other hand,
when the particles become entangled the mode distribution cannot
be factored and, in general, the r. h. s. of the above equation is
not zero \cite{nci}. The state admits the usual statistical
interpretation and the system can be prepared in that state. Note
that this result  is independent of the spin states of the three
particles (of course, the two identical fermions are in the same
one).

One can think that the use of boundless integrations could be
responsible for the appearance of infinite values in the
normalization of these states. In particular, the use of Dirac's
delta in the calculations involving (anti)commutators can be
suspicious in this respect. To discard this possibility, we have
repeated the calculations in a finite volume version of the
problem. The integration is replaced by sums over discrete indexes
related to the quantized momenta of the particles. Now the
creation operator of the composite particle is $\hat{a}_f^+ =\sum
_ {n,m} f_{nm} \hat{a}_n^+ \hat{b}_m^+ $ with $f_{nm}$ the
coefficient of the composite mode $nm$, with normalization $\sum _
{n,m} f_{nm}^* f_{nm}=1$. The expressions for the
(anti)commutators are $[\hat{b}_n , \hat{b}_m^+]=\delta _{nm}$ and
$\{ \hat{a}_n , \hat{a}_m^+ \}=\delta _{nm}$. After a simple
calculation we have $ N_3^{-1}=-1+\sum _{n,m,r} f_{nm}^* f_{rm}
g_r^* g_n$. The same conclusions of the continuous case are valid
for the discrete one. It is simple to verify by direct calculation
that all the other results derived in this paper remain valid,
with obvious modifications, for discrete systems.

The result of this section has been derived in the second
quantization formalism. In the Appendix we show that the same
conclusion is obtained in the more familiar first quantization
formalism.

In conclusion, the example considered here shows that in some
scenarios the existence of entanglement between a fermion and a
distinguishable particle can preclude the application of the
exclusion principle between two identical fermions. Next we shall
study the scope of this result.

\subsection{Scopes of the method and the result}

It is important to correctly understand the scopes of the above
method and its result in order to know when the exclusion principle
can or cannot be applied in the presence of entanglement. The
first limitation comes from the assumption of the system of
entangled particles to be described as a single entity in the
second quantization formalism. As discussed in the previous
section, from a physical point of view, this assumption is not
plausible in some situations. We must analyse in every particular
scenario when it is and when it is not.

The framework considered above excludes the case when the
third particle with which one of the fermions interacts is an
indistinguishable one. The analysis of this problem is much more
involved than that presented here. When the third particle is
indistinguishable of the pair of fermions there are two
types of entanglement, the dynamical entanglement associated with
interactions between the particles and the statistical
entanglement originated with the indistinguishable nature of the
particles. For instance, in Ref. \cite{cor} the existence of entanglement in an ideal (without interaction) gas of
identical fermions has been shown. Then both types of entanglement must be
considered separately, becoming necessary a different type of
analysis.

An example within the range of applicability of our method, but
with an opposite result, is that of several identical fermions
interacting with other distinguishable particles in bound states,
for instance, electrons in atoms. The three-particle case
corresponds to the helium atom (in the approximation that the
nucleus can be treated as a single particle, although it is a
composite one). Now there is simultaneous interaction between the
three particles. The three-particle system can no longer be
decomposed into a free fermion and an entangled two-particle
system. The momentum distribution cannot be separated in the form
$f(p,q)g(P)$. The three-particle state now has the form
\begin{equation}
|\tilde{3}> =N_{\tilde{3}} \int \int \int dp dq dP f(p,q,P) \hat{a}_ {p,r}^+
\hat{a} _{q,s}^+ \hat{b}_{P,S}^+ |0>
\end{equation}
We explicitly include the spin states for the sake of
clarity. The normalization constant is:
\begin{equation}
N_{\tilde{3}}^{-2} = \int \int \int dp dq dP f^* (p,q,P)(f(p,q,P)-f(q,p,P)\delta _{rs})
\end{equation}

For symmetric interactions between the three particles (as in the
case of atoms where all the interactions depend on the positions
of the particles in the form $|{\bf r}_i - {\bf r}_j|$) we have
$f(p,q,P)=f(q,p,P)$, as can be easily seen by transforming the
wavefunction to the momentum representation. Then for fermions in
the same state, $r=s$, we have $ N_{\tilde{3}}^{-2} =0$. The
exclusion principle can be applied to the system although there is
entanglement. On the other hand, when $r \neq s$ we have a finite
normalization constant and the state is physically admissible. We
recover the usual results for these types of systems. We see that
our method to decide if the exclusion principle must be applied to
a particular state, based on the normalization constant of the
state, accounts for three- particle states such as atoms.

To sum up, the analysis presented in this subsection establishes
the scopes of the method and the result presented in the previous
subsection. Our method can only be applied to pairs of fermions
that interact with distinguishable particles. This excludes some
important scenarios as the ideal gas. On the other hand, if the
interaction is with a distinguishable particle our result is valid
when the third particle is only entangled with one of the two
fermions and, in addition, as discussed in the previous section,
the entangled fermion-distinguishable particle system can be
described as a single entity in the second quantization formalism.

\section{Discussion}

We have derived the general (anti)commutation relations of
creation and annihilation operators of composite particles. We
have also studied the role of Pauli's principle in composite
systems. It is exactly valid for "composite fermions" (in spite of
the fact that the complete set of anticommutation relations
differs from that of pure fermions) when the description of the
two interacting particles as a single entity in the second
quantization formalism is physically meaningful. As discussed in
Sect. 4 a good framework to discuss this physical relevance is
that of the creation/annihilation operators. One must decide when
the operators associated with the composite system provide a good
description of the system and when the correct description is
given by the operators of the component particles. For pairs of
particles in bound states it seems intuitive that the composite
system must be described as a composite particle. However, for
other systems as entangled two-particle systems not in bound
states the answer is not clear and must be elucidated for every
particular system.

We have also presented an example of two identical fermions, one
of them entangled with a distinguishable particle, where Pauli's
principle cannot be applied. In our example the temporal sequence
of the interactions is fundamental. The interaction with the
distinguishable particle must be well separated in time from the
overlapping of the two identical fermions (see also discussion of the initial conditions in the Appendix).
This example shows that in these types of situations we must decide
if the exclusion principle can or cannot be applied. The
mathematical technique used in this paper is based on the
normalization constant of the state. We have seen that this
procedure also gives a correct description of atoms with two
electrons or similar systems where Pauli's principle must be
applied. As discussed in Sect. 5.3 the problem of two fermions
interacting with an indistinguishable particle has some
peculiarities due to the simultaneous presence of dynamical and
statistical entanglement. These particular characteristics must be
considered in detail before analysing this type of problems (in particular the relation between entanglement and Pauli's principle)
along the line of this paper.

The inhibition of the exclusion principle in the example discussed
in this paper can be surprising. However, it only reflects a
fundamental property of entangled systems, where the
characteristics of the components of the complete system can be modified.
In general, we cannot characterize them in the same
way as when they are isolated (in non entangled states). In some
circumstances a particle in an entangled state can become distinguishable from
an identical particle, in spite of the fact that in the absence of
entanglement the two particles would be indistinguishable. In
other words, in some circumstances (the helium atom is an example
of a system where this property is not valid) a component of an
entangled system loses partially its identity within the larger
system. A well-known analogy useful to illustrate this point is
the analysis of interference patterns of two-particle systems in
entangled states, where the existence of entanglement prevents the
formation of one-particle interferences (a fundamental characteristic of its behaviour when several alternative paths are available to the particle), which would be present in
the absence of entanglement \cite{Sil}.

The relation between entanglement and Pauli's principle could be
specially relevant for open systems. In open systems, where it is
impossible to completely shield the two identical fermions, almost
certainly any of the fermions will interact with particles of the
environment. We identify, as usual in decoherence theory
\cite{Zeh}, all the particles external to the two fermions with
the environment. If the environment contains particles that become
entangled with any of the fermions the action of the exclusion
principle can be suppressed (of course, in addition to this
entanglement, the rest of conditions under which we have derived
this result must hold). Pauli's principle is potentially fragile
in open systems. The resemblance with decoherence theory is
evident. In decoherence theory the entanglement with the particles
of the environment locally suppresses the interference terms.
Similarly, the exclusion principle can be fragile in the presence of
entanglement with particles of the environment.

Finally, we want to remark that in this paper we have restricted
our discussions to one-dimensional systems. However, it is simple
to verify that all the results derived in the previous sections
remain valid in three-dimensional systems. In this case, the
creation operator of the composite particle is $\hat{c}_f^+ =\int
d^3 {\bf p} \int d^3 {\bf q} f({\bf p}, {\bf q}) \hat{a}_{{\bf
p},r}^+ \hat{b}_{{\bf q},s}^+ $. Using the (anti)commutation
relations $[\hat{a}_{{\bf p},r} \hat{a}_{{\bf q},s}^+ ]=\delta ^3
({\bf p}-{\bf q})\delta _{rs},...$ we can see by simple
calculation that the (anti)commutation relations derived in Sect.
2 stand (with obvious modifications). Similarly, using instead of
Eq. (\ref{eq:nene}) the distribution $f({\bf p},{\bf q}) \sim
exp(-\alpha {\bf p}^2 -\beta {\bf q}^2 -2 \gamma {\bf p}.{\bf q})$
we obtain the same type of relation between expectation values and
entanglement described in Sect. 3. Neither behaviour of the
normalization constant used in Sect. 5 is modified by the
transition to three dimensions. In relation to the
dimensionality of the system an interesting situation is the
two-dimensional case. When the dynamics of some composites (as,
for instance, a charged particle bound to a tube of magnetic flux)
can be confined to two dimensions the system can exhibit
fractional statistics \cite{Wil}. Even in some cases it is
possible to transmute continually the statistics of the system
from the fermionic one to the bosonic one (see, for instance,
\cite{Ved} where the transmutation is obtained by varying the degree
of entanglement between the two subsystems). It would be an
interesting problem to study if the presence of composite
particles would modify the description of these types of
two-dimensional systems.

{\bf Acknowledgments}

This work has been partially supported by the DGICyT of the
Spanish Ministry of Education and Science under contract n.
REN2003-00185 CLI.

{\bf Appendix}

We derive now the results of Sect. 5 in the framework of first
quantization, where the indistinguishable character of the
particles is explicitly taken into account. For any system of
three particles with two of them indistinguishable fermions the total
wavefunction of the complete system must be antisymmetrized with
respect to the variables of the identical particles,
\begin{equation}
\Psi ({\bf x}, {\bf y}, {\bf z}) \rightarrow \Psi ({\bf x}, {\bf y}, {\bf z}) -\Psi ({\bf y}, {\bf x}, {\bf z})
\label{eq:auno}
\end{equation}
where ${\bf x}$ and ${\bf y}$ are the coordinates of the identical
particles and ${\bf z}$ those of the distinguishable one. For
simplicity, we do not explicitly include the normalization
coefficients or the spin variables.

We first consider the case of Sect. 5.2. After the interaction of
the distinguishable particle with one of the identical ones they
become entangled in the form $ \psi ({\bf x}, {\bf z})$; solution
of the equation
\begin{equation}
i\hbar \frac{\partial \psi}{\partial t} =\left( -\frac{\hbar ^2}{2m_I} \nabla _{ x}^2 -\frac{\hbar ^2}{2m_D} \nabla _{ z}^2 +V({\bf x}, {\bf z})   \right) \psi
\end{equation}
with $m_I$ and $m_D$ the masses of the two types of
particles and $V$ the potential ruling the interaction between
them.

On the other hand, the other identical particle obeys the free-particle equation
\begin{equation}
i\hbar \frac{\partial \phi}{\partial t} =-\frac{\hbar ^2}{2m_I} \nabla _{ y}^2  \phi
\end{equation}
Later, when the overlapping of the wavefunctions of the identical
particles is not negligible (for simplicity we assume other
interactions between them to be negligible) Eq. (\ref{eq:auno})
becomes:
\begin{equation}
\psi ({\bf x}, {\bf z}) \phi ({\bf y}) -\psi ({\bf y}, {\bf z}) \phi ({\bf x})
\end{equation}
In general, this expression is different from zero, and the
probability of finding the particles in that state is not null.
However, if $V$ does not entangle the particles the state becomes
($\psi ({\bf x}, {\bf z}) = \phi _* ({\bf x}) \phi _D({\bf z}) $,
where subscripts $*$ and $D$ denote, respectively, that they
are solutions of the same free-equation but with different initial
conditions and mass):
\begin{equation}
\phi _*({\bf x}) \phi _D ({\bf z}) \phi ({\bf y}) -\phi _* ({\bf y}) \phi _D ({\bf z}) \phi ({\bf x})
\end{equation}
When $\phi _* = \phi $ (the identical fermions are in the same
state) the expression becomes null: the probability of finding the
fermions at the same position is zero.

When interaction $V$ entangles the particles in the temporal
order specified here, the exclusion principle does not, in general,
act between the indistinguishable particles.

The situation is completely different in the helium atom (as an
example of Sect. 5.3). The ruling equation is now
\begin{equation}
i\hbar \frac{\partial \Psi}{\partial t} =\left( -\frac{\hbar
^2}{2m_I} (\nabla _{x}^2 + \nabla _{y}^2   )-\frac{\hbar ^2}{2m_D}
\nabla _{z}^2 +U({\bf x}, {\bf y}, {\bf z})   \right) \Psi
\end{equation}
Taking into account that all the terms are invariant under the
interchange $ {\bf x} \leftrightarrow {\bf y}$ (the Coulomb
potential depends on $|{\bf x}-{\bf y}|$, just as the
spin-spin interaction when it is taken into account), the
Hamiltonian is invariant. Thus, if the initial conditions are also
invariant under the interchange, $\Psi _0({\bf x}, {\bf y}, {\bf
z}) = \Psi _0({\bf y}, {\bf x}, {\bf z})$ (note that in particular
the two spins must be equal), we have $\Psi ({\bf x}, {\bf y},
{\bf z}) = \Psi ({\bf y}, {\bf x}, {\bf z})$ and the configuration
is excluded by the exclusion principle.

The key role of the initial conditions in the
above reasoning must be remarked. In the case of Sect. 5.2 there is a delay between
the entangling interaction and the overlapping of the identical
fermions, and the initial wavefunction is not invariant under the
exchange $ {\bf x} \leftrightarrow {\bf y}$.

We finally note that we have obtained the same result in the first
quantization framework by antisymmetrizing the wavefunction and in

the second quantization one without antisymmetrizing the relevant
states. This is so because in the second case the effects of
antisymmetrization are generated by the anticommutation relations
(see Ref. \cite{Shi} for similar considerations in a different
context).


\begin{thebibliography}{99}
\bibitem{nci} This statement can be explicitly demonstrated
in the case of Gaussian distributions. For the distributions
(\ref{eq:nene}) and $g(p)=(2 \alpha /\pi)^{1/4}exp(-\alpha p^2)$
the integral gives $\sqrt{2}(\alpha \beta - \gamma ^2
)^{1/2}(2\alpha \beta -\gamma ^2 )^{-1/2}$, expression that only
equals $1$ for $\gamma =0$.
\end{thebibliography}
\end{document}